\documentclass[twoside,twocolumn,english,aps,showpacs,prl,superscriptaddress]{revtex4-1}

	\usepackage[usenames,dvipsnames]{xcolor}
	\usepackage{amsmath}
	\usepackage{amsfonts}
	\usepackage{amssymb}
	\usepackage[colorlinks=true,citecolor=blue,linkcolor=red]{hyperref}
	\usepackage{graphicx}
	\usepackage{bbold}					
	\usepackage[makeroom]{cancel}		
	\usepackage{multirow}				
	\usepackage[normalem]{ulem}        
	\usepackage{array}
	\usepackage{comment}
	\usepackage{pdfpages}
	\makeatletter
	\AtBeginDocument{\let\LS@rot\@undefined}
	\makeatother
	
	\newcolumntype{x}[1]{>{\centering\let\newline\\\arraybackslash\hspace{0pt}}p{#1}}

	\newcounter{subeqn} %
	\makeatletter
	\@addtoreset{subeqn}{equation}
	\makeatother

\usepackage{dcolumn}  

\newtheorem{theorem}{Theorem}

\newcommand{\Eq}[2][Eq.~]{#1(\ref{eq:#2})}
\newcommand{\Fig}[2][Fig.~]{#1\ref{fig:#2}}

\newcommand{\theo}[2][Theorem~]{#1\ref{theorem:#2}}

\newcommand{\expect}[1]{{\left\langle{#1}\right\rangle}}

\newcommand{\map}[1]{{\mathcal{#1}}}
\newcommand{\nmap}[1]{\widetilde{\mathcal{#1}}}

\begin{document}

\title{Noise-resilient phase estimation with randomized compiling}

\author{Yanwu Gu}
\email{guyw@baqis.ac.cn}
\affiliation{Beijing Academy of Quantum Information Sciences, Beijing 100193, China}
\affiliation{State Key Laboratory of Low Dimensional Quantum Physics, Department of Physics, Tsinghua University, Beijing, 100084, China}

\author{Yunheng Ma}
\affiliation{Beijing Academy of Quantum Information Sciences, Beijing 100193, China}
\affiliation{State Key Laboratory of Low Dimensional Quantum Physics, Department of Physics, Tsinghua University, Beijing, 100084, China}

\author{Nicol\`o Forcellini}
\affiliation{Beijing Academy of Quantum Information Sciences, Beijing 100193, China}

\author{Dong E. Liu}
\email{dongeliu@mail.tsinghua.edu.cn}
\affiliation{State Key Laboratory of Low Dimensional Quantum Physics, Department of Physics, Tsinghua University, Beijing, 100084, China}
\affiliation{Beijing Academy of Quantum Information Sciences, Beijing 100193, China}
\affiliation{Frontier Science Center for Quantum Information, Beijing 100184, China}
\affiliation{Hefei National Laboratory, Hefei 230088, China}

\date{\today}

\begin{abstract}
We develop an error mitigation method for the control-free phase estimation. We prove a theorem that under the first-order correction, the phases of a unitary operator are immune to the noise channels with only Hermitian Kraus operators, and therefore, certain benign types of noise for phase estimation are identified. By further incorporating the randomized compiling protocol, we can convert the generic noise in the phase estimation circuits into stochastic Pauli noise, which satisfies the condition of our theorem. Thus we achieve a noise-resilient phase estimation without any quantum resource overhead. The simulated experiments show that our method can significantly reduce the estimation error of the phases by up to two orders of magnitude. Our method paves the way for the utilization of quantum phase estimation before the advent of fault-tolerant quantum computers.
\end{abstract}

\maketitle

\textbf{{\em Introduction.}}--- Quantum phase estimation (QPE) \cite{nielsen&chuang, cleve1998quantum} is a crucial component in quantum algorithms which are believed to achieve an exponential speedup over their classical counterparts for solving problems, such as integer factorization \cite{shor1994algorithms}, linear systems of equations \cite{harrow2009quantum}, the Hamiltonian spectrum \cite{abrams1999quantum}. 
However, the conventional phase estimation algorithm, which is based on the quantum Fourier transform (QFT), requires ancillary qubits to perform the controlled unitary operators and quantum error correction (QEC)~\cite{ShorIEEE1996,Aharonov-threshold,Preskill1998,Knill1998,kitaev,fowler} to combat noise. These requirements surpass the capability of the current noisy intermediate-scale quantum (NISQ) devices \cite{preskill2018quantum,aruteSupremacy,wu2021strong,Pino-TrapIon2021-Science}. 


Regarding the controlled unitary operations in the conventional schemes, their implementation typically necessitates the employment of many more native gates than the unitary itself.
Some variants of the phase estimation algorithm have been devised to reduce the number of control qubits to one \cite{kitaev1995quantum,aspuru2005simulated,knill2007optimal,higgins2007entanglement,svore2013faster,wiebe2016efficient,o2019quantum,somma2019quantum}. Furthermore, the control-free phase estimation \cite{kimmel2015robust,roushan2017spectroscopic,russo2021evaluating,neill2021accurately,lu2021algorithms,zintchenko2016,o2021error,lin2022heisenberg,Dutkiewicz2022heisenberglimited} is a more appealing variant in the NISQ era, where the unitary of interest is repeatedly applied instead of its control version. 
Control-free phase estimation was first applied to calibrate the unitary errors in single-qubit gates as ``robust phase estimation'' \cite{kimmel2015robust}, and later adapted to two-qubit gate calibration as ``Floquet calibration'' \cite{neill2021accurately}.


Noise affects all quantum operations in NISQ devices. Understanding and correcting the effect of noise is necessary for meaningful quantum computations. In the NISQ era, quantum error mitigation (QEM) \cite{o2021error,kandala2019error,temme2017error,li2017efficient,temme2017error,endo2018practical,mcclean2017hybrid,bonet2018low,sagastizabal2019experimental,mcardle2019error,mcclean2020decoding,arute2020observation,mi2021information,mi2022time,montanaro2021error} is a more feasible technique than the resource-expensive QEC.  Among the QEM schemes, verified phase estimation (VPE) \cite{o2021error} is a technique that is directly linked to the QPE. This method achieves error mitigation by postselecting the shots in which the system register is recovered to its initial state. Due to the postselection, each circuit requires many extra repetitions to collect enough statistics. The number of repetitions is inversely proportional to the circuit fidelity, which decreases exponentially as the number of gate operations increases. In addition, VPE cannot mitigate the unitary noise, which is one of the most common types of noise in multi-qubit circuits, due to ineluctable miscalibrations and crosstalks \cite{Sarovar2020detectingcrosstalk}. As a result, it is urgently necessary to develop a more practical scheme for general noise kinds that uses less resources in NISQ devices.




In this work, we develop an error mitigation technique for control-free phase estimation using randomized compiling, with no resource overhead other than some efficient classical computations. We theoretically analyze the noise types that the phase estimation is insensitive to. We prove a theorem that if all Kraus operators of the associated noise channel are Hermitian, 
the phases extracted from noisy QPE circuits do not change under the first-order correction in comparison to those from ideal circuits. In addition, the randomized compiling method is applied to convert general noise types including coherent noise in the circuit to stochastic Pauli noise, which fits the criterion in our theorem. As a result, we achieve a practical error reduction in phase estimation. The validity of our error mitigation method is tested using two simulated experiments. The results show that our method can reduce errors in both unitary and stochastic noise, particularly in unitary noise by up to two orders of magnitude.

\begin{figure}
    \centering
    \includegraphics[width=\linewidth]{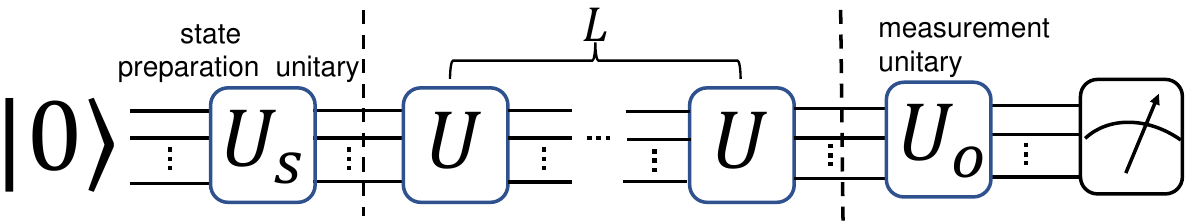}
    \caption{The circuit structure of control-free phase estimation. The initial state is prepared by applying a unitary operator $U_s$. The unitary $U$ is then repeated for $L$ times. Finally, the expectation value of an operator $O$ is measured by performing a measurement unitary $U_{O}$ prior to the computational basis measurement.}
    \label{fig:control free circuit}
\end{figure}

\textbf{{\em Control-free phase estimation.}}--- Let us briefly review the control-free phase estimation \cite{kimmel2015robust,roushan2017spectroscopic,russo2021evaluating, neill2021accurately} where its circuit structure is shown in \Fig{control free circuit}. The spectral decomposition of the unitary operator $U$ is
\begin{equation} \label{eq:unitary operator}
    U= \sum_{a} e^{i \lambda_a} |\phi_a\rangle \langle \phi_a|
\end{equation}
where $\lambda_a$ is a phase and $|\phi_a\rangle$ is 
the corresponding eigenstate. The system is prepared to a state $|\psi\rangle =\sum_{a}c_a |\phi_a\rangle$ by conducting a unitary operator $U_s$. To obtain the phases of interest, the initial state $|\psi\rangle$ should be carefully chosen so that it only has a few relevant eigenstates of $U$. The unitary operator $U$ is then repeatedly applied to $|\psi\rangle$ for $L$ times, where $L$ is an integer ranging from $1$ to $L_{\textrm{max}}$. Finally, an operator $O$ is measured by performing a unitary $U_{O}$ before the computational basis measurement. The expectation value of $O$ is 
\begin{eqnarray}
    \expect{O}_L &=& \langle \psi| (U^\dagger)^L O U^L  |\psi\rangle  \nonumber\\
    &=& \sum_{a,b} c_a c^*_{b} \langle \phi_b|O|\phi_a\rangle e^{ i (\lambda_a-\lambda_b) L} \,.
\end{eqnarray}
The difference $\lambda_a-\lambda_b$ can be retrieved by conducting a discrete Fourier transform on the vector of $\expect{O}_L$ or by performing a function fitting. Control-free phase estimation, like randomized benchmarking \cite{knill2008randomized,magesan2011scalable}, is robust to state preparation and measurement (SPAM) errors, because SPAM errors only affect the coefficients $c_ac_b^*\langle \phi_b|O|\phi_a\rangle$ but not phases (this is also true for other phase estimation algorithms with repeated control version of target unitary).

To get the individual phases, we assume a reference eigenstate $|\phi_0\rangle$ exists with known phase $\lambda_0$. Then, we prepare an initial state $|\psi\rangle = \frac{1}{\sqrt{2}} \left( |\phi_0\rangle + |t\rangle \right)$, where $|t\rangle =\sum_{n=1}^{N_p} c_n |\phi_n\rangle $ is a state containing $N_p$ eigenstates whose phases are to be estimated. The measurement operator is chosen as non-Hermitian $O=2|\phi_0\rangle \langle t|$, which can be always decomposed as the sum of Hermitian operators~\cite{nielsen&chuang}. In this case, the expectation value reads 
\begin{eqnarray} \label{eq:exp ideal}
 \expect{O}_L&=& 2\langle \psi| (U^\dagger)^L   |\phi_0\rangle \langle t| U^L |\psi\rangle \nonumber\\
 &=& \sum_{n}c_n c^{*}_n e^{ i (\lambda_n-\lambda_0)L} \,.
\end{eqnarray}
Finally, the measurements of the decomposed Hermitian operators yield the phases $\{\lambda_n\}$.


 


\textbf{{\em Benign type of noise for phase estimation.}}---
Any quantum algorithms running on the current devices are affected by noise. If we can identify the noise types that are mild to phase estimation, we may tailor the noise in quantum devices to the desired type and achieve error mitigation.

We use the language of quantum channels (represented with Calligraphical symbols) to describe the noise effect. The corresponding unitary channel $\mathcal{U}$ of the unitary operator $U$ in \Eq{unitary operator} has the effect
\begin{equation}
    \mathcal{U}\left( |\phi_a\rangle \langle \phi_b| \right) = U |\phi_a\rangle \langle \phi_b| U^\dagger = e^{ i (\lambda_a-\lambda_b)} |\phi_a\rangle \langle \phi_b|\,.
\end{equation}
Thus $|\phi_a\rangle \langle\phi_b|$ is an eigen-operator of the unitary channel $\mathcal{U}$ with eigenvalue $e^{ i (\lambda_a-\lambda_b)}$. If a noisy channel $\mathcal{E}(\rho)=\sum_{k}E_k \rho E_{k}^{\dagger}$ is appended to the unitary channel $\mathcal{U}$, then the resultant noisy gate is $\widetilde{\mathcal{U}}=\mathcal{E}\, \mathcal{U}$. In Sec.II of supplementary information (SI)~\cite{supp}, we prove that the expectation value in \Eq{exp ideal} is transformed into 
\begin{equation} \label{eq:exp noisy}
    \expect{O}_{L} \approx  \sum_{n}p_n \, (g_{n0})^L\, e^{ i \lambda_{n0}L}
\end{equation}
under weak noise $\map{E}$, where $p_n \approx c_n c_n^{*}$ is a real number close to the proportion of $|\phi_n\rangle$ in $|t\rangle$. The eigenvalue of the noisy operation $\nmap{U}$, that is deviated from the ideal unitary operation $\map{U}$, is modified to $g_{n0}e^{ i \lambda_{n0}}$ with the noisy amplitude $g_{n0}$ and phase $\lambda_{n0}$, respectively. 
The amplitude has the constraint $g_{n0}\le 1$ for completely positive trace preserving (CPTP) maps describing physical channels \cite{QChannelLecture.pdf, watrous2018theory}. Thus \Eq{exp noisy} is a damping oscillating model. One can still conduct a Fourier transform or fit \Eq{exp noisy} to obtain an estimation of the noisy phases $\lambda_{n0}$.  To reduce the inaccuracy in the phase estimation, we can identify noise types benign for the phase estimation, and then design a scheme to convert the general noise to those desired types. Following this strategy, we first provide a theorem below.
\begin{theorem}
\label{theorem:noise condition}
If every Kraus operator $E_k$ of a noise $\mathcal{E}$ is Hermitian, then the noisy version $\widetilde{\mathcal{U}}=\mathcal{E}\,\mathcal{U}$ of a unitary channel
$\mathcal{U}$ keeps the phases unchanged up to the leading order of noise strength.
\end{theorem}
This theorem is based on the first-order perturbation theory (see Sec.I of SI~\cite{supp} for the detailed proof, which includes Refs.~\cite{sakurai1995modern,kato2013perturbation,sarkar1995,potts2013,helsen2019spectral,oBrien_2019}). The proof makes no assumptions regarding the form of the unitary operator $U$, and therefore, the theorem applies to any phase estimation problem. Our theorem provides a sufficient condition regarding the benign types of noise, which already encompasses a variety of noise types, including stochastic Pauli noise. Particularly, the phase damping channel ($T_2$ error) is a stochastic Pauli channel.

The noise in a real device is exceedingly complex and rarely meets the criterion in \theo{noise condition}. For example, the amplitude damping channel ($T_1$ error) and the unitary noise cannot satisfy the condition of our theorem. Fortunately, there is a technique known as randomized compiling (RC) \cite{wallman2016noise,hashim2021randomized} that can turn the noise in the corresponding circuit of $\mathcal{U}$ into stochastic Pauli noise. Note RC can only convert the noise in each cycle of $U$ to stochastic Pauli noise, but we suppose the noise channel $\map{E}$ occurs at the end of $U$ in \theo{noise condition}. As a result, the condition in \theo{noise condition} is stronger than what RC can achieve; see the detailed discussion below.



\textbf{{\em The effect of randomized compiling.}}---
Randomized compiling (RC) \cite{wallman2016noise,hashim2021randomized}  is a technique to reduce the general and complex noise in circuits to a specific simple noise type, namely stochastic Pauli noise. Initially, the bare circuit $U$ is partitioned into many cycles and each cycle contains a layer of single-qubit gates and a layer of two-qubit gates, that is 
\begin{equation}
    U=G_K C_K \cdots G_k C_k \cdots G_1 C_1
\end{equation}
 where $C_k$ and $G_k$ represent single-qubit and two-qubit layers, respectively. Each layer of single-qubit
gates $C_k$ is replaced with a round
of randomized dressed gates $\widetilde{C}_k = T_k C_k T_{k-1}^{c}$ where $T_k$ are chosen uniformly at random from the Pauli group and the correction operators are set to $T_{k}^{c}= G_k T_{k}^{\dagger} G_{k}^{\dagger}$~\cite{wallman2016noise,hashim2021randomized}. In the RC protocol, each dressed gate layer $\widetilde{C}_k$ should be implemented as a single layer of elementary gates. Thus, randomized compiling can create a set of randomized circuits that are logically equivalent to the original circuit without increasing circuit depth. The noise in circuits is customized to stochastic Pauli noise after averaging the outputs of these randomized circuits. 


For phase estimation in \Fig{control free circuit}, we perform RC on the circuit segments $U^L$. Running these randomized circuits with the same initial state and final measurement, and taking the average of results tailor the noise in each cycle to stochastic Pauli noise. Now the actual operations by the circuit $U^L$ can be written as 
\begin{equation}
       \left(\mathcal{G}_K \mathcal{E}_{p}^{(K)} \mathcal{C}_K \cdots \mathcal{G}_1 \mathcal{E}_{p}^{(1)} \mathcal{C}_1 \right)^L
        = \left(\mathcal{E'}\, \mathcal{U} \right)^L
\end{equation}
where $\mathcal{E}_p$ is a stochastic Pauli channel. The noise $\mathcal{E}'$ is defined by moving all the $\mathcal{E}_p$s to the end of each segment $U$. We ask whether the composite noise channel $\map{E}'$ after RC fits the condition in \theo{noise condition}?


If $U$ is a Clifford circuit, the final error is still a Pauli error, which meets our criterion. However, if the circuit $U$ includes some non-Clifford gates, the condition in \theo{noise condition} is generally not satisfied. For the stochastic Pauli noise after the RC protocol, we can perform a probabilistic analysis. First, if we assume the probability of error in each cycle is relatively small, there is a high possibility that only one Pauli error (denoted as $P$) occurs in $U$. We assume the unitary operator of the circuit segments before and after the error $P$ to be $U_1$ and $U_2$, and thus the ideal unitary is $U= U_2 U_1$. With error $P$, the whole operation becomes $U_2 P U_1$. Let us move $P$ to the end of $U$, and obtain  $U_2 P U_2^\dagger U_2U_1 = (U_2 P U_2^\dagger) U$. Now, the new error operator $U_2 P U_{2}^{\dagger}$ is Hermitian, which clearly satisfies the condition of \theo{noise condition}. For the cases of two errors and beyond, many cases still satisfy the criteria in \theo{noise condition}, for example, if the circuit segments between 
the first $P_{1}$ and the last $P_{n}$  error are all Clifford.



Now, we provide more rigorous results for the effect of RC (see detailed analysis in Sec.III of SI~\cite{supp}).
We show that the phase error in bare circuits is proportional to the $\|\mathcal{E}-\mathcal{I}\|_{\diamond}$, the diamond norm distance between the noise $\map{E}$ in bare circuit and identity channel $\map{I}$. After RC, the phase error is proportional to the $\|\mathcal{E}'-\mathcal{I}\|_{\diamond}^\alpha$ where $1<\alpha \le 2$. For Clifford circuit $U$, $\alpha =2$; for the non-Clifford circuit, $\alpha < 2$ because the criterion in \theo{noise condition} can only partially be met. If the gate noise in bare circuit $U$ is stochastic with a characteristic noise probability $p$, the phase error is $\sim p$ in the bare circuits and $\sim p^\alpha$ in the RC circuits. For the case of unitary noise with the characteristic rotation angle $\theta$, the 
phase error in the bare circuits is $\sim \theta$. After RC, the unitary error is converted into stochastic Pauli noise with some noise probability $p\sim \theta^2$ \cite{kliesch2021theory}. The phase error by RC circuits should be $\sim p^\alpha \sim \theta^{2\alpha}$. These results show that our method can reduce the estimation error of the phases for both stochastic and unitary noise, particularly has a stronger effect on unitary noise. Note the theoretical scaling may not be obtained in actual experiments due to the finite number of shots, random circuits, and repetitions of $U$. Nevertheless, our method shows strong error mitigation power under practical experimental settings as shown by the simulated experiments.

\textbf{{\em Simulated experiments.}}---To demonstrate the performance of our method, we present two simulated experiments: 1) The estimation of the quasi-energies of a Floquet system~\cite{neill2021accurately}, 2) an order finding problem \cite{nielsen&chuang}. The measured signals $\expect{X}_L + i \expect{Y}_L$ are first Fourier transformed to the frequency domain, and then the peak locations in the frequency spectrum provide a rough estimate of phases. To obtain a more accurate estimate, one can fit the data to the expression in \Eq{exp noisy}. The estimation error is defined as the average distance between the estimates $\widehat{\lambda}_n$ using our method and the actual phases $\lambda_n$, i.e., $\frac{1}{N}\sum_{n=1}^N \left |\widehat{\lambda}_n-\lambda_n \right |$ where $N$ is the number of non-degenerate actual phases. See Sec. IV of SI~\cite{supp} for the circuits and more simulation details.

\begin{figure}
    \centering
    \includegraphics[width=1\columnwidth]{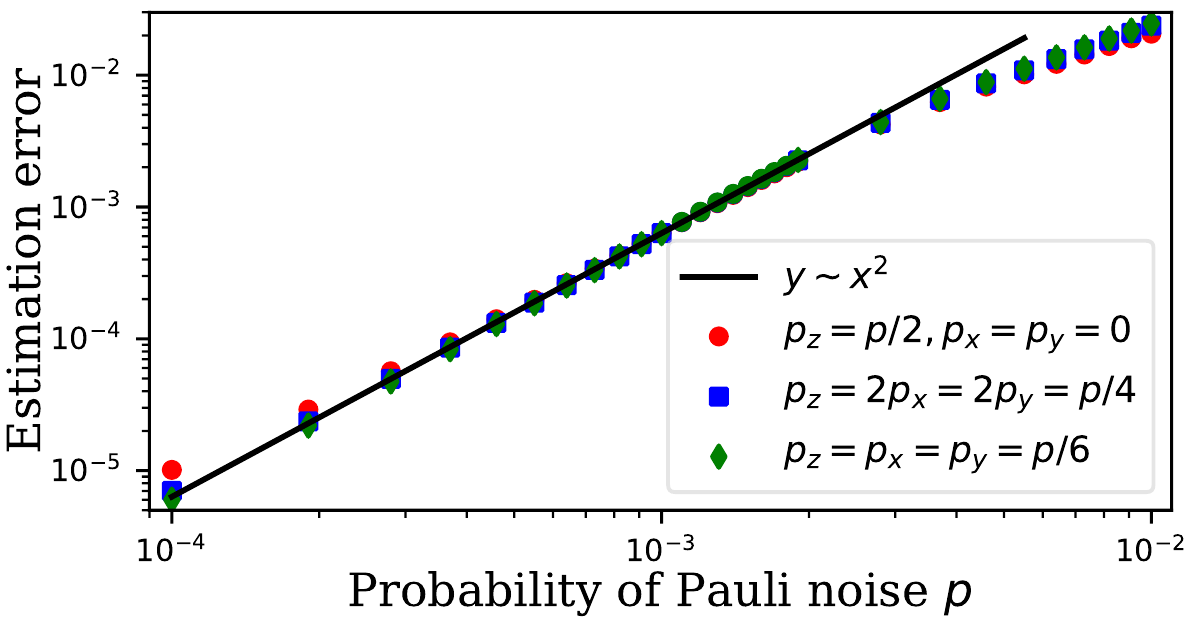}
    \caption{Phase estimation error under stochastic Pauli noise. Each qubit at each gate location is subjected to a stochastic Pauli channel with Pauli noise probabilities $p_x,p_y,p_z$. Under very weak noise, the scaling power approaches 2. But for relatively large noise, the scaling power is less than 2 because the non-negligible cases of two or more Pauli errors in circuit $U$ may violate the condition in \theo{noise condition}.}
    \label{fig:ideal_scaling}
\end{figure}

Firstly, we use a 6-qubit version of the Floquet system in Ref.~\cite{neill2021accurately} to validate the phase estimation error scaling with the noise strength after RC. But this scaling can only be obtained in a very ideal case, i.e., an infinite number of shots $N_s$ (to eliminate sampling error), an infinite number of random circuits $N_r$ for each bare circuit (to perfectly transform noise into stochastic Pauli noise) and enough number of repetitions $L$ of target unitary (to remove the possible effect of not enough data). Thus, we directly compute the expectation values from the final density matrix of each circuit. Furthermore, we set the noise of each gate to be stochastic Pauli noise and use a very large circuit length $L_{\textrm{max}}$. As shown in \Fig{ideal_scaling}, when the noise probability of each gate is very low, the probability of two or more Pauli errors occurring in the circuit $U$ is negligible, and thus the scaling power $\alpha$ approaches 2. However, when noise probability is relatively large, the cases of two and more Pauli errors cause $\alpha$ to be less than 2, which coincides with our theoretical analysis.

Then, we show the performance of our method under practical experimental settings, i.e., a finite number of $N_s, N_r$ and a small $L_{\textrm{max}}$. We add unitary noise or stochastic noise (with non-Hermitian Kraus operators) to each gate. The shot for each bare circuit is set to $N_s$. We use RC to generate $N_r=20$ randomized circuits for each bare circuit. To maintain the same resource cost, we set the shot for each randomized circuit to $N_s/N_r$. 

\begin{figure}[t!]
    \centerline{\includegraphics[width=1\columnwidth]{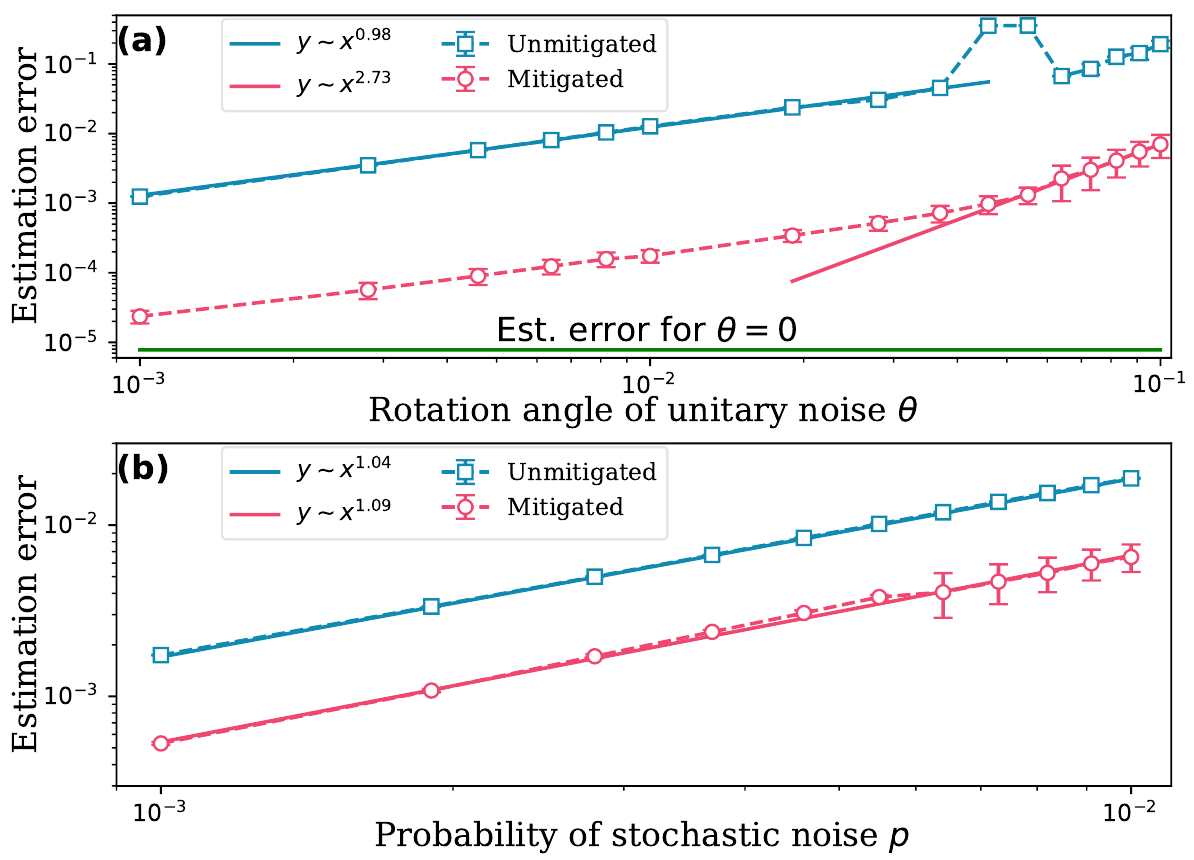}}
    \caption{Simulated results on estimating Floquet quasi-energies \cite{neill2021accurately} ($10$-qubit system), under unitary noise (a) and stochastic noise (b). At small $\theta$, the estimation error decreases slowly due to a small number of $N_r$. One can increase $N_r$ to get more error mitigation power because the phase estimation error scales with $N_r$ as $1/\sqrt{N_r}$, see Sec.~IV of SI~\cite{supp}.}
	\label{fig:google sim results}
\end{figure}



In \Fig{google sim results}, we show the results of simulated experiments on estimating Floquet quasi-energies with $N_s=10^7,L_{\textrm{max}}=50$. The unitary operation $U$ is a 10-qubit non-Clifford unitary operator.  As shown in \Fig{google sim results}(a), our technique has strong error mitigation power for unitary noise. The estimation error is reduced by up to two orders of magnitude. If there is no error mitigation, it is linearly increased with the rotation angle of unitary noise. For the error mitigated case in \Fig{google sim results}(a), we fit the results in the strong noise regime with a power law function and obtain its power of almost $2.73$. Both the scaling behaviors of estimation error with or without error mitigation are consistent with our analytical analysis. The absence of theoretical scaling behaviour in the weak noise regime is a result of the small number of random circuits $N_r$.  In Sec. IV of SI \cite{supp}, we show the phase estimation error scales with $N_r$ as $1/\sqrt{N_r}$.  In actual experiments, $N_r$ can be increased up to $N_s$ in order to increase error mitigation capability without incurring experimental overhead. Though not as significant as in the case of unitary noise, our 
method provides a nearly $60\%$ reduction in estimation error for stochastic noise as shown in \Fig{google sim results}(b).

\begin{figure}[t!]
    \centerline{\includegraphics[width=1\columnwidth]{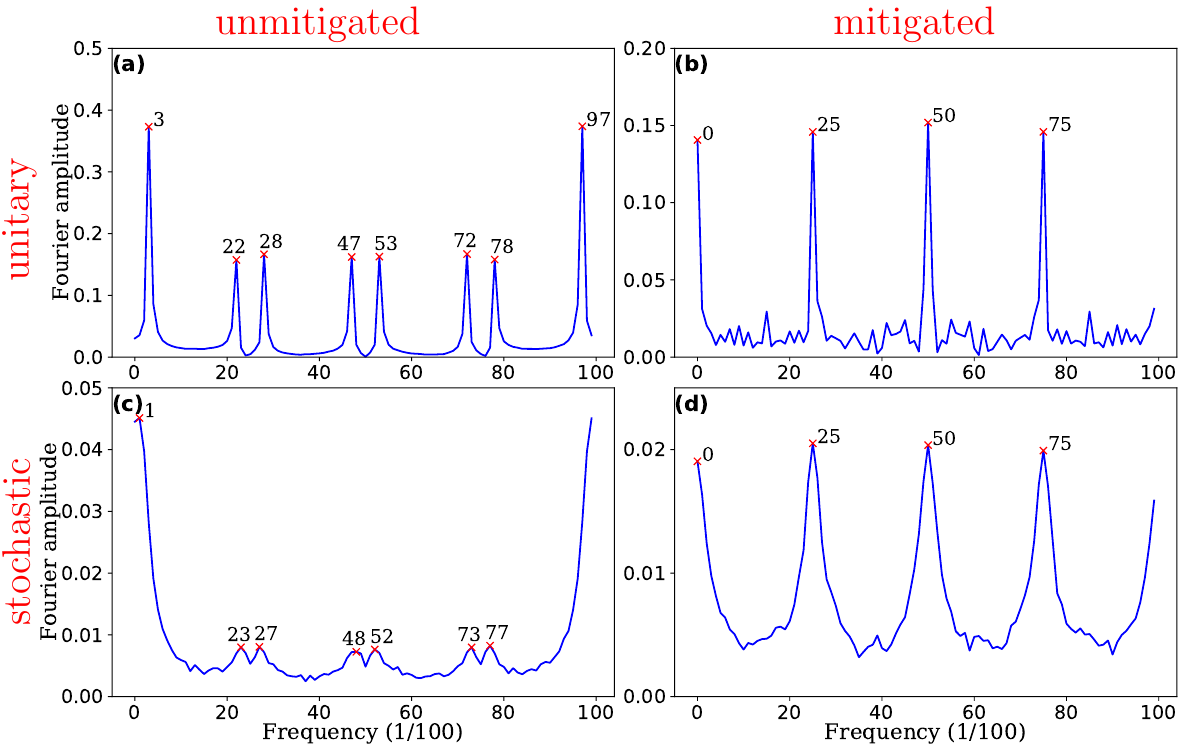}}
    \caption{Simulated results of order finding $x=4$ and $N=255$, for two types of noise: unitary noise (a)(b) and stochastic noise (c)(d). The correct peaks in the frequency domain are located at 0, 25, 50, 75. Some extra peaks appear in (a)(c) due to the splitting of degenerated eigenvalues under noise. Our method completely eliminates these extra peaks and the remaining peaks are located at the correct positions in (b)(d). 
    }
    \label{fig:of sim results}
\end{figure}

Finally, we consider the order finding problem, that is to find the least positive integer $r$ such that for two specified coprime numbers $x$ and $N$ $(x< N)$ we have $x^r=1 (\mathrm{mod}\,N)$. Here we simulate an order finding problem with $x=4$ and $N=255$. The corresponding unitary is an 8-qubit Clifford unitary with only \textrm{SWAP} gates~\cite{supp}. The main numerical results are shown in \Fig{of sim results} (with $N_s=10^5$), where we present the Fourier transform of the measured signals under unitary noise $(a,b)$ and stochastic noise $(c,d)$. The maximum number of repetitions of the unitary of interest is $L_{\textrm{max}}=100$. In this particular order finding problem,  the order should be $r=4$, which induces four peaks at 0, 25, 50, 75 in the frequency domain. Under unitary noise, degenerated eigenvalues split, resulting in extra peaks in \Fig{of sim results}(a). With our error mitigation method, these extra peaks are completely removed and the surviving peaks are located at the correct positions in \Fig{of sim results}(b). Under stochastic noise, some peaks are almost buried by noise in \Fig{of sim results}(c). However, our method recovers the right peaks from noise in \Fig{of sim results}(d).


\textbf{{\em Conclusion}}---
We have developed an efficient method to mitigate the errors in control-free phase estimation. The method is based on \theo{noise condition} that the noise with Hermitian Kraus operators causes only mild changes in the phases of a unitary operator. To achieve the desired type of noise in actual devices, randomized compiling can be used, which has no resource overhead.  The two simulated experiments on Hamiltonian eigenvalue estimation and order finding demonstrate the strong error mitigation power of our method. We emphasize that our scheme requires an assumption that the noise is time independent, and this assumption is reasonable, at least in the ``Good devices''~\cite{aruteSupremacy,wu2021strong,neill2021accurately}. In addition, the effect of the noise drift can be controlled and calibrated, for example, using the method~\cite{proctor2020detecting}. Finally, our method can be readily applied to the phase estimation scheme with a single control qubit.




\begin{acknowledgements}
\textbf{\em Acknowledgments.}---
This work was supported by the Innovation Program for Quantum Science and Technology (Grant No. 2021ZD0302400), the National Natural Science Foundation of China (Grant No. 12147123 and 11974198), and the Beijing Natural Science Foundation (No. Z220002). The source code for the simulated experiments is available at this site \url{https://github.com/yanwu-gu/noise-resilient-phase-estimation}.
\end{acknowledgements}

\bibliography{reference}
\bibliographystyle{apsrev4-1}

\onecolumngrid
\newpage


\section*{Supplementary material}

Yanwu Gu,[1,2] Yunheng Ma,[1,2] Nicolò Forcellini,[1] and Dong E. Liu[2,1,3,4]

[1]*Beijing Academy of Quantum Information Sciences, Beijing 100193, China*
[2]*State Key Laboratory of Low Dimensional Quantum Physics,
Department of Physics, Tsinghua University, Beijing, 100084, China*
[3]*Frontier Science Center for Quantum Information, Beijing 100184, China*
[4]*Hefei National Laboratory, Hefei 230088, China*
(Dated: June 20, 2023)

In this supplementary, we provide some details about the central results in the main body of the paper.

- Sec. I provides a detailed proof of Theorem 1 by using the first-order perturbation theory.

- Sec. II derives the fitting function used in the phase estimation with noise.

- Sec. III presents an analytical analysis of our method's error mitigation power and the uncertainty of the phase estimation by functional fit.

- Sec. IV shows some details about the simulated experiments, including the circuits used.

## I. PROOF OF THEOREM 1 OF THE MAIN TEXT

With phase estimation, we want to identify the eigenvalues of a unitary operator $U$, which has eigen-decomposition

$$U|\phi_a\rangle = e^{i\lambda_a}|\phi_a\rangle. \tag{S1}$$

As the circuit implementing $U$ is inevitably associated with some noise, it is more convenient to use quantum channels rather than quantum operators. Quantum channels are completely-positive trace-preserving (CPTP) maps, which transform one operator to another. The action of a quantum channel $\mathcal{E}$ on an arbitrary operator $O$ can be characterized by a set of Kraus operators $E_k$, i.e., $\mathcal{E}(O) = \sum_k E_k O E_k^\dagger$. We denote the corresponding unitary channel of the unitary operator $U$ as $\mathcal{U}$, whose action on an operator $O$ is $\mathcal{U}(O) = UOU^\dagger$. Thus, the unitary channel $\mathcal{U}$ has the eigen-decomposition

$$\mathcal{U}(|\phi_a\rangle\langle\phi_b|) = U|\phi_a\rangle\langle\phi_b|U^\dagger = e^{i(\lambda_a - \lambda_b)}|\phi_a\rangle\langle\phi_b|. \tag{S2}$$

Quantum channels are linear maps that can be represented as matrices under a set of the basis operators of the operator space, such as eigen-operators $|\phi_a\rangle\langle\phi_b|$. Meanwhile, operators are represented as vectors. The associated inner product between two operators $A$ and $B$ is the Hilbert-Schmidt inner product $\text{tr}\{A^\dagger B\}$. Therefore, in this representation $\mathcal{U}$ is a unitary matrix.

Let us append a noise channel $\mathcal{E}$ to $\mathcal{U}$, with the noisy version of $\mathcal{U}$ denoted as $\widetilde{\mathcal{U}} = \mathcal{E}\mathcal{U}$. We investigate the relationship between the eigenvalues of the $\widetilde{\mathcal{U}}$ and those of $\mathcal{U}$. If the noise is relatively weak, the problem is an eigenvalue perturbation of the unitary matrix [1]. Given the close relationship between unitary and Hermitian matrices, one can use Hermitian matrix perturbation theory [1, 2] to get the correction of eigenvalues and eigenstates, assuming a diagonalizable noisy gate $\widetilde{\mathcal{U}}$. In most cases, the assumption should be met in actual devices, since diagonalizable matrices are dense in the space of all matrices, meaning that any non-diagonalizable matrix can be deformed into a diagonalizable one by a small perturbation. In the following, we apply Hermitian perturbation theory to obtain the first order correction of the eigenvalues and hence prove Theorem 1.

*Proof.* The perturbation matrix is

$$\Delta = \widetilde{\mathcal{U}} - \mathcal{U} = (\mathcal{E} - \mathcal{I})\mathcal{U}. \tag{S3}$$

We assume the perturbation is small in terms of some norm, such as the diamond norm $\|\Delta\|_\diamond = \delta$ [3]. Then, for a non-degenerate

eigenvalue $e^{i(\lambda_a - \lambda_b)}$ with eigen-operator $|\phi_a\rangle\langle\phi_b|$, the first order correction is

$$
\begin{aligned}
\epsilon &= \mathrm{tr}\left\{(|\phi_a\rangle\langle\phi_b|)^\dagger \Delta(|\phi_a\rangle\langle\phi_b|)\right\} \\
&= \mathrm{tr}\left\{|\phi_b\rangle\langle\phi_a|(\mathcal{E}-\mathcal{I})\mathcal{U}(|\phi_a\rangle\langle\phi_b|)\right\} \\
&= e^{i(\lambda_a-\lambda_b)}\left[\mathrm{tr}\{|\phi_b\rangle\langle\phi_a|\mathcal{E}(|\phi_a\rangle\langle\phi_b|)\}-1\right] \\
&= e^{i(\lambda_a-\lambda_b)}\left[\mathrm{tr}\left\{|\phi_b\rangle\langle\phi_a|\sum_k E_k|\phi_a\rangle\langle\phi_b|E_k^\dagger\right\}-1\right] \\
&= e^{i(\lambda_a-\lambda_b)}\left[\sum_k\langle\phi_a|E_k|\phi_a\rangle\langle\phi_b|E_k^\dagger|\phi_b\rangle-1\right].
\end{aligned}
\tag{S4}
$$

If every Kraus operator $E_k$ of $\mathcal{E}$ is Hermitian, i.e., $E_k = E_k^\dagger$, the correction $\epsilon$ alters only the amplitude of the eigenvalue but not the phase.

For degenerate eigenvalues $e^{i\lambda_n}$ with eigen-operators $|\phi_a\rangle\langle\phi_b|$ satisfying $\lambda_a - \lambda_b = \lambda_n$, these eigen-operators span a subspace. The $ab, a'b'$-entry of the perturbation matrix in this subspace is

$$
\begin{aligned}
\Delta_{ab,a'b'} &= \mathrm{tr}\left\{(|\phi_a\rangle\langle\phi_b|)^\dagger(\mathcal{E}-\mathcal{I})\mathcal{U}(|\phi_{a'}\rangle\langle\phi_{b'}|)\right\} \\
&= e^{i\lambda_n}\left[|\phi_b\rangle\langle\phi_a|\sum_k E_k|\phi_{a'}\rangle\langle\phi_{b'}|E_k^\dagger-\delta_{aa'}\delta_{bb'}\right] \\
&= e^{i\lambda_n}\left[\sum_k\langle\phi_a|E_k|\phi_{a'}\rangle\langle\phi_{b'}|E_k^\dagger|\phi_b\rangle-\delta_{aa'}\delta_{bb'}\right] \\
&= e^{i\lambda_n}V_{ab,a'b'}
\end{aligned}
\tag{S5}
$$

where we introduced a new matrix $V$ whose entry is $V_{ab,a'b'} = \sum_k\langle\phi_a|E_k|\phi_{a'}\rangle\langle\phi_{b'}|E_k^\dagger|\phi_b\rangle - \delta_{aa'}\delta_{bb'}$. It's easy to prove that the matrix $V$ is Hermitian if $E_k = E_k^\dagger$. The $a'b', ab$-entry of $V$ is

$$
\begin{aligned}
V_{a'b',ab} &= \sum_k\langle\phi_{a'}|E_k|\phi_a\rangle\langle\phi_b|E_k^\dagger|\phi_{b'}\rangle - \delta_{aa'}\delta_{bb'} \\
&= \left(\sum_k\langle\phi_a|E_k^\dagger|\phi_{a'}\rangle\langle\phi_{b'}|E_k|\phi_b\rangle\right)^* - \delta_{aa'}\delta_{bb'} \\
&= \left(\sum_k\langle\phi_a|E_k|\phi_{a'}\rangle\langle\phi_{b'}|E_k^\dagger|\phi_b\rangle - \delta_{aa'}\delta_{bb'}\right)^* \\
&= V_{ab,a'b'}^* = V_{a'b',ab}^\dagger.
\end{aligned}
\tag{S6}
$$

Thus the matrix $V$ is Hermitian, and its eigenvalues are real. The eigenvalues of $\Delta$ in the degenerated subspace are the product of $e^{i\lambda_n}$ with reals. Therefore, the first order correction to the eigenvalue only changes the amplitude and keeps the phase. $\qquad\square$

Based on the Theorem 3.9 of Kato book [1], we give a sufficient condition for the convergence of the perturbation series. For a particular eigenvalue $e^{i(\lambda_a - \lambda_b)}$, we define its isolation distance $r_{ab}$ as the smallest distance between the eigenvalue $e^{i(\lambda_a - \lambda_b)}$ and other unequal eigenvalues, that is

$$
r_{ab} = \min_{\lambda_c - \lambda_d \neq \lambda_a - \lambda_b}|e^{i(\lambda_a - \lambda_b)} - e^{i(\lambda_c - \lambda_d)}|.
\tag{S7}
$$

Then, a sufficient condition on the noise strength to ensure the perturbation series for the eigenvalue $e^{i(\lambda_a - \lambda_b)}$ convergent is that

$$
\|\Delta\|_1 = \|(\mathcal{E}-\mathcal{I})\mathcal{U}\|_1 = \|\mathcal{E}-\mathcal{I}\|_1 < \frac{1}{2}r_{ab}
\tag{S8}
$$

where $\|\Delta\|_1$ is the induced trace norm of the channel $\Delta$ and defined as $\|\Delta\|_1 = \max_{X;\|X\|_1\leq 1}\|\Delta(X)\|_1$ ($\|\Delta(X)\|_1$ is the trace norm of the operator $\Delta(X)$).

## II. THE FITTING FUNCTION

In this section, we derive the fitting function for phase estimation with noise. We assume noisy gate $\widetilde{\mathcal{U}}$ has eigen-operators $M_{ab}$, that is

$$\widetilde{\mathcal{U}}(M_{ab}) = g_{ab} e^{i\lambda_{ab}} M_{ab} \tag{S9}$$

where $g_{ab}$, $\lambda_{ab}$ are the amplitude and phase of the eigenvalue, respectively. For any CPTP map, we have $0 \leq g_{ab} \leq 1$ [4, 5]. Eigenvalues and eigen-operators always come in conjugate pairs, since

$$
\begin{aligned}
\widetilde{\mathcal{U}}(M_{ab}^{\dagger}) &= \sum_k F_k M_{ab}^{\dagger} F_k^{\dagger} \\
&= \sum_k \left( F_k M_{ab} F_k^{\dagger} \right)^{\dagger} \\
&= \left( \widetilde{\mathcal{U}}(M_{ab}) \right)^{\dagger} \\
&= g_{ab} e^{-i\lambda_{ab}} M_{ab}^{\dagger},
\end{aligned}
\tag{S10}
$$

where $F_k$ are Kraus operators of $\widetilde{\mathcal{U}}$.

Assuming that $\widetilde{\mathcal{U}}$ is diagonalizable, then the $M_{ab}$ forms a basis of the operator space. But these basis operators $M_{ab}$ may not be orthonormal. It's convenient to introduce the corresponding left eigen-operators $G_{ab}$ of $M_{ab}$ when using $M_{ab}$ as a basis to expand other operators. They satisfy the relation

$$\mathrm{tr}\left\{ G_{a'b'}^{\dagger} M_{ab} \right\} = \delta_{a'b',ab} \,. \tag{S11}$$

If the noise is weak, $M_{ab}, G_{ab}$ can be expressed as

$$M_{ab} = M_{ab}^0 + \sum_{cd \neq ab} h_{cd}^{ab} M_{cd}^0; \quad G_{ab} = G_{ab}^0 + \sum_{cd \neq ab} s_{cd}^{ab} G_{cd}^0 \tag{S12}$$

where $M_{ab}^0 (G_{ab}^0)$ is the unperturbed (left) eigen-operator with eigenvalue $e^{i(\lambda_a - \lambda_b)}$. $h_{cd}^{ab} (s_{cd}^{ab})$ is the coefficient on the basis $M_{cd}^0 (G_{cd}^0)$ in the correction of the eigen-operator $M_{ab}^0 (G_{ab}^0)$. The first-order correction has a norm of the order $\|\mathcal{E} - \mathcal{I}\|_{\diamond} = \delta$. For non-degenerate eigenvalues, $M_{ab}^0 = G_{ab}^0 = |\phi_a\rangle\langle\phi_b|$; for degenerate eigenvalues, $M_{ab}^0, G_{ab}^0$ are linear superposition of the degenerate eigen-operators $|\phi_a\rangle\langle\phi_b|$ of the ideal unitary channel $\mathcal{U}$. $M_{ba}$ is defined as the perturbed operator of $M_{ba}^0$ with eigenvalue $e^{i(\lambda_b - \lambda_a)}$ which is the unperturbed eigenvalue of $M_{ab}^{\dagger}$. Thus we get

$$M_{ab}^{\dagger} = M_{ba} \,. \tag{S13}$$

Similarly, we have

$$G_{ab}^{\dagger} = G_{ba} \,. \tag{S14}$$

The noisy initial state $\widetilde{\rho}$ can be expanded as

$$\widetilde{\rho} = \sum_{ab} \mathrm{tr}\left\{ G_{ab}^{\dagger} \widetilde{\rho} \right\} M_{ab} \,. \tag{S15}$$

After $L$ repetitions of the noisy gate $\widetilde{\mathcal{U}}$, the expectation value of the noisy operator $\widetilde{O}$ is

$$
\begin{aligned}
\left\langle \widetilde{O} \right\rangle_L &= \mathrm{tr}\left\{ \widetilde{O} \widetilde{\mathcal{U}}^L(\widetilde{\rho}) \right\} \\
&= \sum_{ab} \mathrm{tr}\left\{ \widetilde{O} M_{ab} \right\} \mathrm{tr}\left\{ G_{ab}^{\dagger} \widetilde{\rho} \right\} g_{ab}^L e^{i\lambda_{ab} L} \\
&= \sum_{a \leq b} g_{ab}^L \left[ \mathrm{tr}\left\{ \widetilde{O} M_{ab} \right\} \mathrm{tr}\left\{ G_{ab}^{\dagger} \widetilde{\rho} \right\} e^{i\lambda_{ab} L} + \mathrm{tr}\left\{ \widetilde{O} M_{ab}^{\dagger} \right\} \mathrm{tr}\left\{ G_{ab} \widetilde{\rho} \right\} e^{-i\lambda_{ab} L} \right] \,.
\end{aligned}
\tag{S16}
$$

If $\widetilde{O}$ is a Hermitian operator, the expectation value can be further simplified

$$\left\langle \widetilde{O} \right\rangle_L = \sum_{a \leq b} g_{ab}^L (C_{ab} e^{i\lambda_{ab}L} + C_{ab}^* e^{-i\lambda_{ab}L}) \tag{S17}$$

where $C_{ab} = \mathrm{tr}\left\{\widetilde{O} M_{ab}\right\} \mathrm{tr}\left\{G_{ab}^\dagger \widetilde{\rho}\right\}$. This function is always real as expected. The range of $a, b$ is determined by the number of non-trivial eigen-operators in the initial state $\widetilde{\rho}$ and final operator $\widetilde{O}$. Here we can only estimate the differences between phases. In some problems, this information is enough, such as the order finding problem where the differences take the same values as the phases themselves.

To estimate the absolute values of phases, one needs to prepare a state $|\psi\rangle = \frac{1}{\sqrt{2}}(|\phi_0\rangle + |t\rangle)$ where $|t\rangle = \sum_{n=1}^{N_p} c_n |\phi_n\rangle$. $N_p$ is the number of eigenstates in $|t\rangle$. The operator $O = 2|\phi_0\rangle\langle t|$ is measured at the end. If the SPAM errors are small, we have $\widetilde{\rho} \approx \rho = |\psi\rangle\langle\psi|$ and $\widetilde{O} \approx O$. By Eq. (S16), we obtain

$$\begin{aligned}
\left\langle \widetilde{O} \right\rangle_L &= 2\sum_n c_n^* \sum_{ab} \mathrm{tr}\{|\phi_0\rangle\langle\phi_n|M_{ab}\}\mathrm{tr}\left\{G_{ab}^\dagger \rho\right\} g_{ab}^L e^{i\lambda_{ab}L} \\
&\approx 2\sum_n c_n^* \sum_{ab} \mathrm{tr}\{|\phi_0\rangle\langle\phi_n|M_{ab}^0\}\mathrm{tr}\left\{G_{ab}^{0\dagger} \rho\right\} g_{ab}^L e^{i\lambda_{ab}L}
\end{aligned} \tag{S18}$$

where we just keep the zeroth order term of $\delta$ in the $M_{ab}$ and $G_{ab}$. The effect of the first order correction will be discussed later. The operator $|\phi_0\rangle\langle\phi_n|$ acts as a filter, that is, for a fixed $n$, only $M_{ab}^0$ with the same eigenvalue as that of $|\phi_n\rangle\langle\phi_0|$ have nonzero factors. For a non-degenerate operator $|\phi_n\rangle\langle\phi_0|$, $M_{ab}^0$ must be equal to $|\phi_n\rangle\langle\phi_0|$, which contributes a term $c_n c_n^* g_{n0}^L e^{i\lambda_{n0}L}$ in $\left\langle \widetilde{O} \right\rangle_L$. For degenerate $|\phi_n\rangle\langle\phi_0|$, all the $M_{ab}^0$ in the degenerate subspace have nonzero factors. Their corresponding perturbed eigenvalues $g_{ab} e^{i\lambda_{ab}}$ become different due to noise. That is, there are more oscillating terms than in the ideal expectation value, resulting in more peaks in the frequency domain. However, if the noise is sufficiently weak, all of the associated perturbed eigenvalues $g_{ab} e^{i\lambda_{ab}}$ are still very close. Thus, we can remove the dependence of the eigenvalues on the index $a, b$ and replace them with $g_{n0} e^{i\lambda_{n0}}$. The expectation value is simplified as

$$\begin{aligned}
\left\langle \widetilde{O} \right\rangle_L &\approx 2\sum_n c_n^* \mathrm{tr}\left\{\sum_{ab} \mathrm{tr}\{|\phi_0\rangle\langle\phi_n|M_{ab}^0\} G_{ab}^{0\dagger} \rho\right\} g_{n0}^L e^{i\lambda_{n0}L} \\
&= 2\sum_n c_n^* \mathrm{tr}\{|\phi_0\rangle\langle\phi_n|\rho\} g_{n0}^L e^{i\lambda_{n0}L} \\
&= \sum_n c_n^* c_n \, g_{n0}^L \, e^{i\lambda_{n0}L} \\
&= \sum_n p_n \, g_{n0}^L \, e^{i\lambda_{n0}L}
\end{aligned} \tag{S19}$$

where $p_n = c_n c_n^*$. The second line holds because $G_{ab}^{0\dagger} = G_{ba}^0$ is a basis of the degenerate subspace where $|\phi_0\rangle\langle\phi_n|$ lives. This is the fitting function we use in the simulated experiments. The term in the first order correction, which causes an oscillation other than $\lambda_{n0}$, is

$$\sum_n c_n^* \sum_{ab} \mathrm{tr}\left\{|\phi_0\rangle\langle\phi_n| \sum_{cd \neq ab} h_{cd}^{ab} M_{cd}^0\right\} \mathrm{tr}\left\{G_{ab}^{0\dagger} \rho\right\} g_{ab}^L e^{i\lambda_{ab}L}.$$

The amplitude of each such damping oscillation mode is

$$\begin{aligned}
p_{ab} &= \left|\sum_n c_n^* \mathrm{tr}\left\{|\phi_0\rangle\langle\phi_n| \sum_{cd \neq ab} h_{cd}^{ab} M_{cd}^0\right\} \mathrm{tr}\left\{G_{ab}^{0\dagger} \rho\right\}\right| \\
&\leq \sum_n |c_n^*| \left|\mathrm{tr}\left\{|\phi_0\rangle\langle\phi_n| \sum_{cd \neq ab} h_{cd}^{ab} M_{cd}^0\right\}\right| \left|\mathrm{tr}\left\{G_{ab}^{0\dagger} \rho\right\}\right| \\
&\approx N_p \frac{1}{\sqrt{N_p}} \delta = \sqrt{N_p} \delta
\end{aligned} \tag{S20}$$

where each $c_n$ is assumed to be in the order $1/\sqrt{N_p}$. The first order correction in $\text{tr}\{G_{ab}^\dagger \rho\}$ just slightly changes the factor $p_n$ without introducing further oscillations. Thus, to prevent incorrect phases from being estimated, we require $p_{ab} \ll p_n$, that is

$$\delta \ll \frac{1}{N_p^{3/2}} \,. \tag{S21}$$

## III. THE ACCURACY AND PRECISION OF PHASE ESTIMATION

This section examines our method's error mitigation capability as well as the uncertainty of the phase estimate.

We initially investigate the error for phase estimation in the absence of randomized compiling. From Eq. (S4) we get the first order correction to the eigenvalue

$$\begin{aligned}
\epsilon &= e^{i(\lambda_a - \lambda_b)}\text{tr}\{(|\phi_a\rangle\langle\phi_b|)^\dagger(\mathcal{E} - \mathcal{I})(|\phi_a\rangle\langle\phi_b|)\} \\
&= e^{i(\lambda_a - \lambda_b)}f_{ab}e^{i\,d_{ab}}
\end{aligned} \tag{S22}$$

where we rewrite the expression using the noisy amplitude $f_{ab}$ and the phase $d_{ab}$: $\text{tr}\{(|\phi_a\rangle\langle\phi_b|)^\dagger(\mathcal{E} - \mathcal{I})(|\phi_a\rangle\langle\phi_b|)\} = f_{ab}e^{i\,d_{ab}}$. It's easy to prove that

$$f_{ab} \leq \|\mathcal{E} - \mathcal{I}\|_\diamond \tag{S23}$$

where $\|\mathcal{E} - \mathcal{I}\|_\diamond$ is the diamond norm. From Hölder's inequality, we have

$$\begin{aligned}
f_{ab} &= \left|\text{tr}\{(|\phi_a\rangle\langle\phi_b|)^\dagger(\mathcal{E} - \mathcal{I})(|\phi_a\rangle\langle\phi_b|)\}\right| \\
&\leq \||\phi_a\rangle\langle\phi_b|\|_\infty \|(\mathcal{E} - \mathcal{I})(|\phi_a\rangle\langle\phi_b|)\|_1 \\
&= \|(\mathcal{E} - \mathcal{I})(|\phi_a\rangle\langle\phi_b|)\|_1 \\
&\leq \max_{X:\|X\|_1 \leq 1}\|(\mathcal{E} - \mathcal{I})(X)\|_1 \\
&= \|\mathcal{E} - \mathcal{I}\|_1
\end{aligned} \tag{S24}$$

where $\|X\|_p = \text{tr}\left\{\left(\sqrt{X^\dagger X}\right)^p\right\}^{1/p}$ is the Schatten $p$-norm of the operator $X$ and $\|\mathcal{E} - \mathcal{I}\|_1$ is the induced trace norm of channel $\mathcal{E} - \mathcal{I}$. The diamond norm $\|\mathcal{E} - \mathcal{I}\|_\diamond$ is an induced trace norm on an extendable Hilbert space, hence we get $f_{ab} \leq \|\mathcal{E} - \mathcal{I}\|_1 \leq \|\mathcal{E} - \mathcal{I}\|_\diamond$. We use the diamond norm to bound the magnitude of the first order correction because it is a more commonly used metric for measuring noise strength. It is, nevertheless, a worst-case metric. In some instances, there may be a large discrepancy between $f_{ab}$ and $\|\mathcal{E} - \mathcal{I}\|_\diamond$.

The eigenvalue after the first order correction is

$$\begin{aligned}
e^{i(\lambda_a - \lambda_b)}(1 + f_{ab}e^{i\,d_{ab}}) &= e^{i(\lambda_a - \lambda_b)}(1 + f_{ab}\cos d_{ab} + if_{ab}\sin d_{ab}) \\
&= e^{i(\lambda_a - \lambda_b)}\sqrt{(1 + f_{ab}\cos d_{ab})^2 + (f_{ab}\sin d_{ab})^2}\,e^{i\,h_{ab}}.
\end{aligned} \tag{S25}$$

Thus the phase error $h_{ab}$ satisfy $\tan h_{ab} = \frac{f_{ab}\sin d_{ab}}{1 + f_{ab}\cos d_{ab}}$. We get $|h_{ab}| \sim f_{ab} \leq \|\mathcal{E} - \mathcal{I}\|_\diamond$. This is just the estimation error of a phase, i.e., the difference between the noisy phase and the ideal phase, based on the use of the bare circuits. After randomized compiling (RC) is applied, the noise channel $\mathcal{E}$ in bare circuits is turned into noise channel $\mathcal{E}'$. We know that the first order corrections of eigenvalues induced by $\mathcal{E}'$ do not change phases. Thus, we anticipate the error on a phase in RC circuits is $\sim \|\mathcal{E}' - \mathcal{I}\|_\diamond^\alpha$ where $1 < \alpha \leq 2$. For Clifford circuits, $\alpha = 2$. But for non-Clifford circuits, our method can only partially mitigate errors such that $\alpha < 2$, because the case of two or more errors occurring in circuits could violate Theorem 1.

The reason that Theorem 1 is not fully satisfied is based on the fact: given two error channels $\mathcal{E}_1$ and $\mathcal{E}_2$ that have a representation with Hermitian Kraus operators, their product $\mathcal{E}_2\mathcal{E}_1$ does not necessarily have such a representation. Consider the Kraus operators of the two channels $\mathcal{E}_1, \mathcal{E}_2$ are $E_k^1, E_k^2$, where all the Kraus operators $E_k^1, E_k^2$ are Hermitian. Take the product of any two Kraus operators in the two channels $E_m^2 E_k^1$. The Hermitian conjugate of this product is $E_k^{1\dagger} E_m^{2\dagger} = E_k^1 E_m^2$. Thus, we need to require $E_k^1 E_m^2 = e^{i\lambda}E_m^2 E_k^1$ if we want to have a set of Hermitian Kraus operators (the complex number $e^{i\lambda}$ can be canceled due to the conjugate operation of Kraus operators). But this condition does not hold generally. A simple counter-example is for the two Hermitian operators

$$H = \frac{1}{\sqrt{2}}\begin{pmatrix} 1 & 1 \\ 1 & -1 \end{pmatrix}, \quad Z = \begin{pmatrix} 1 & 0 \\ 0 & -1 \end{pmatrix} \tag{S26}$$

where we have

$$HZ = \frac{1}{\sqrt{2}} \begin{pmatrix} 1 & -1 \\ 1 & 1 \end{pmatrix}, \quad ZH = \frac{1}{\sqrt{2}} \begin{pmatrix} 1 & 1 \\ -1 & 1 \end{pmatrix}. \tag{S27}$$

$HZ$ and $ZH$ can not be equal up to a complex number.

For the cases of two Pauli errors and beyond occurring in $U$, some cases still satisfy the criteria in Theorem 1, for example, if the circuit segments between the first $P_1$ and the last $P_n$ error are all Clifford, i.e. $U_2 P_n Cl_{n-1} P_{n-1} \cdots Cl_1 P_1 U_1$ with Clifford $Cl_i$. Of this situation, one can combine all those errors together by switching error operations with Clifford gates. During the movement of error operations, the Clifford gates transform Pauli errors to other Pauli errors $\tilde{P}_i$. Finally, when the combined error $\tilde{P} = P_n \tilde{P}_{n-1} \cdots \tilde{P}_1$ (still Pauli error) is moved to the end of $U$, the new error operator $U_2 \tilde{P} U_2^\dagger$ is still Hermitian.

To measure the error mitigation power of our method, we should determine the relationship between the two diamond norm distances $\|\mathcal{E} - \mathcal{I}\|_\diamond, \|\mathcal{E}' - \mathcal{I}\|_\diamond$. If the gate noise in bare circuit $U$ is stochastic with a characteristic probability $p$, the diamond norm distance $\|\mathcal{E} - \mathcal{I}\|_\diamond \sim p$. Even though the Kraus operators of gate noise are transformed into Pauli operators after RC, the noise probability should be similar to that in a bare circuit, that is, $\|\mathcal{E}' - \mathcal{I}\|_\diamond \sim p$. Therefore, for stochastic noise, the phase error is $\sim p$ by bare circuits and $\sim p^\alpha$ by RC circuits. However, the diamond distance $\|\mathcal{E}' - \mathcal{I}\|_\diamond$ may differ greatly from $\|\mathcal{E} - \mathcal{I}\|_\diamond$ if gate noise is unitary. Assume the gate noise in the bare circuit is unitary noise with some characteristic rotation angle $\theta$. Typically, the diamond norm distance of $\mathcal{E}$ is $\|\mathcal{E} - \mathcal{I}\|_\diamond \sim \theta$. Thus, the phase error in the bare circuit is $\sim \theta$. After RC, the unitary error is converted into stochastic Pauli noise with some noise probability $p \sim \theta^2$ [3]. The phase error by RC circuits should be $\sim p^\alpha \sim \theta^{2\alpha}$. Thus, our method can achieve very pronounced error mitigation on unitary noise. But due to the limited number of shots, random circuits, and repetitions of target unitary, the theoretical scaling behavior derived here may not be attained practically as shown in Fig. 2(a).

The uncertainty of control-free phase estimation has been thoroughly studied in the previous literature [6, 7]. It is also proved that this approach could achieve the so-called Heisenberg scaling [8, 9]. Following the analysis in the Appendix D of Ref. [7], we give the uncertainty in estimating a phase $\lambda_n$ by fitting the function Eq. (S19)

$$\sigma(\lambda_n) \propto \frac{1}{\sqrt{N_s}} \frac{N_p}{L_{\max}^{3/2}} \tag{S28}$$

where $N_s$ is the number of shots for the circuit in each length. The precision increases as the maximum length $L_{\max}$ increases until

$$L_{\max} \log \frac{1}{g_n} \leq 4 \tag{S29}$$

where $g_n$ is the amplitude of the eigenvalue. For pure unitary noise, $g_n = 1$. For stochastic noise, $g_n \sim 1 - \delta$, which sets an upper bound for the maximum length of the circuits, that is $L_{\max} \leq 4/\delta$.

# IV. DETAILS OF SIMULATED EXPERIMENTS

## A. Estimate of Floquet quasi-energies

The first simulated experiment comes from Ref. [7], where they compute the electronic properties of a quantum ring. The circuits consist of many repetitions of the same unitary cycle whose eigenvalues are to be estimated. The cycle unitary contains two layers of $\sqrt{\text{iSWAP}}$ gates acting on the nearest neighbor qubits (in our simulation, we do not add any disorder, i.e., single-qubit $Z$ rotations). Thus the studied unitary conserves the excitation in the system. The system is first prepared as a superposition state of zero and one excitation by applying a Hadamard gate on the first qubit. Finally, the first qubit is measured on the Pauli-X and Pauli-Y. Because this problem is periodic in both space and time, the eigenstates of the cycle unitary $U$ have well-defined momentum and energies. The eigenvalues $\lambda_k$ of the cycle unitary are

$$\cos \lambda_k = \sin^2(k/2) \tag{S30}$$

where $k$ is the momentum with units of $4\pi/N_q$, starting from 0. $N_q$ is the number of qubits. The original experiment in Ref. [7] involves $N_q = 18$ qubits. In our simulation, we take $N_q = 10$. In our method, we need to perform randomized compiling to mitigate errors, which requires that the two-qubit gates in the circuits are Clifford gates. We thus compile the original circuit with $\sqrt{\text{iSWAP}}$ gates to the logically equivalent circuits but with CNOT gates as two-qubit gates. An example of the compiled circuits with $L = 1$ and Pauli-X measurement is shown in Fig. S1. The number of repetitions of the studied unitary $U$ is up to the $L_{\max} = 50$ in our simulations.

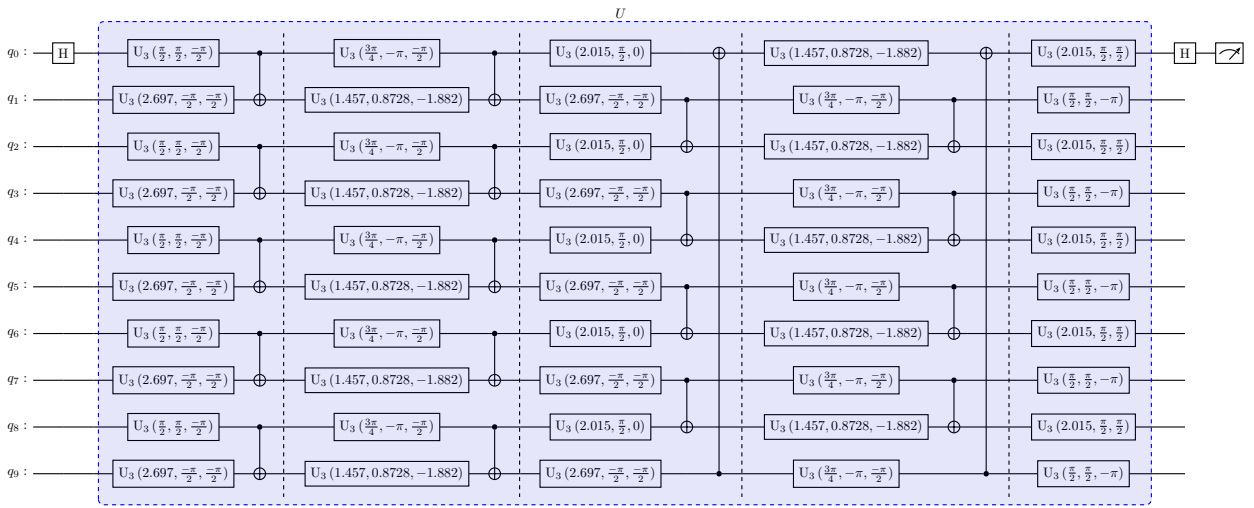

FIG. S1: An example of the circuits used in the simulation of a Floquet quasi-energies estimating problem. Here we study a unitary with $N_q = 10$ qubits. In this example, $L = 1$ and the measurement operator is Pauli-X. The circuit of the studied unitary is the shaded segment between the two Hadamard gates, which contains some non-Clifford single-qubit gates.

We study the effects of two types of noise. One is unitary noise. For each single-qubit gate, we add a $Z$ rotation error $e^{-i\theta_1/2Z}$; and for each two-qubit gate, we add a $Z \otimes Z$ rotation error $e^{-i\theta_2/2Z \otimes Z}$. We set $\theta_2 = 2\theta_1 = \theta$ such that the noise strength of single-qubit gates are almost half of the two-qubit gates ($\theta$ is the rotation angle reported in the main text). Another type is stochastic noise. For each single-qubit gate, we add a phase gate $|0\rangle\langle0| + i|1\rangle\langle1|$ with a probability $p/2$ ($p$ is the noise probability reported in the main text). For each two-qubit gate, we add a gate $e^{-i\pi/4Z \otimes Z}$ with a probability $p$.

Each of the bare circuits is measured $N_s = 10^7$ times. To maintain the same resource cost, we run each random circuit generated from a bare circuit $10^7/N_r$ times, where $N_r = 20$ is the number of random circuits for each bare circuit. We first perform the classical Fourier transform on the data to get a rough estimate of the phases. To obtain a more accurate estimate, one needs to fit the function Eq. (S19). The estimate from the Fourier transform is used as a starting point for the least-square optimization. We measure the estimation error by the average distance between the estimates of functional fit and actual values. The fitting function has the same number of oscillating modes as the ideal case, which can be inferred in this problem. In the strong noise regime, there may be more peaks in the Fourier transform of the data from bare circuits, causing the phase estimation to fail. In the reported results in Fig. 3, the number of peaks in the frequency domain for randomized circuits is always the same as for the ideal case.

### 1. Simulations with $N_s = 10^5$ shots

Here we test the performance of our method under a more practical number of shots $N_s = 10^5$. As shown in Fig. S2, the results are quite similar to the Fig. 3 in the main text except in the regime of weak unitary noise in Fig. S2(a), where the estimation error plateaus due to the limited statistics, i.e., the number of shots $N_s$ (so, this is not an essential limitation of our scheme). The bump at the strong noise part is a consequence of a poor starting point in the least square optimization. Here we use the estimate from the Fourier transform as the starting point for the optimization. The amplitudes of wrong peaks in the bump happen to be larger than those of the correct peaks as shown in Fig. S3(a). In the strong noise regime, the spectrum results without error mitigation contain non-trivial erroneous phases, and therefore, the phase estimation becomes totally unreliable and even meaningless. However, these erroneous phases (or peaks) are completely suppressed by our method as shown in Fig. S3(b).

### 2. Data processing with matrix pencil method

Here, we use another sophisticated signal processing method, the matrix pencil (MP) method [10–13], to extract phases from measured signals. The matrix pencil method includes a singular value decomposition (svd) procedure that allows us to eliminate the oscillating modes with small Fourier amplitudes caused by noise. As shown in Fig. S4(a), the MP method removes the bump

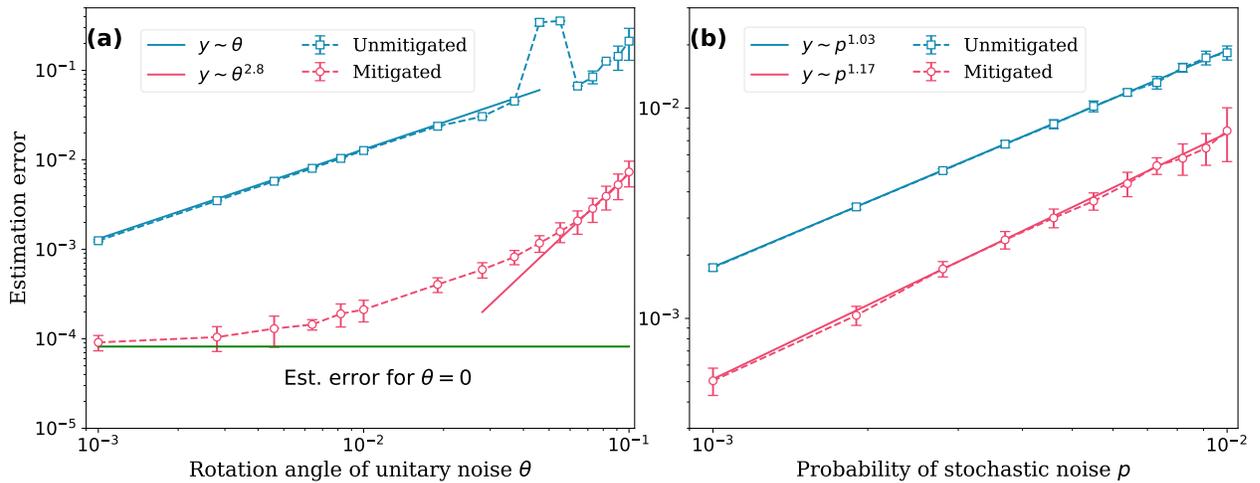

FIG. S2: Simulated results on estimating Floquet quasi-energies [7] (10-qubit system), under unitary noise (a) and stochastic noise (b) with $N_s = 10^5$. At small $\theta$, the estimation error plateaus due to the limited statistics, which is already approaching the estimation error of $\theta = 0$.

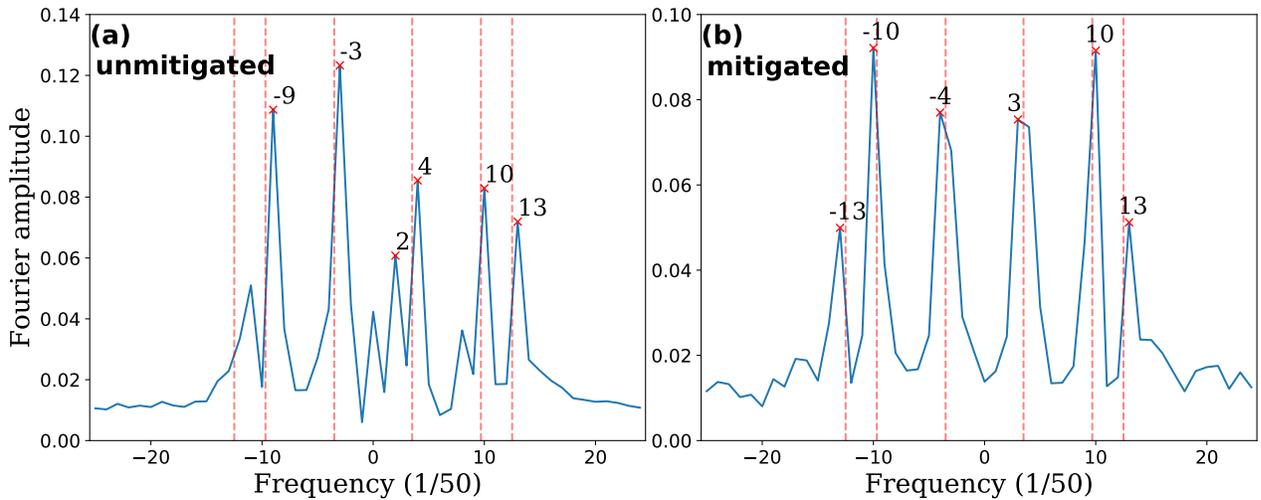

FIG. S3: The Fourier spectrum of the data from the second point of the bump in Fig. S2(a), with error unmitigated (a) and mitigated (b), respectively. Actual frequencies are represented by red dashed vertical lines. The red crosses represent Fourier transform estimations. Without error mitigation, the estimates are far from the actual values, resulting in phase estimation failure.

in the unmitigated case when unitary noise is large. It seems that the svd procedure in MP methods discards the wrong peaks in Fig. S3(a). But how this svd procedure affects phase estimation is not clear, as the estimations from functional fit outperform those from the MP method for most noise parameters.

### 3. The scaling of estimation error with the number of random circuits $N_r$

Here, we show how the phase estimation error trends with the number of random circuits $N_r$ in RC with two simulated experiments. We set noise as a unitary error with angle $\theta = 0.01$. For a bare circuit, we generate $N_r$ random circuits, where each random circuit is run $N_s/N_r$ times to keep the resource cost unchanged. We first perform a simulation with a practical number of shots $N_s = 10^5$. As shown in Fig. S5(a), the estimation error converges very fast with the number of circuits $N_r$, which is consistent with the finding in Ref. [14]. The converged value is in fact the limit set by the sampling error. We want to study how the phase estimation error scales with $N_r$ before it reaches some limits (due to the finite number of shots or the error mitigation

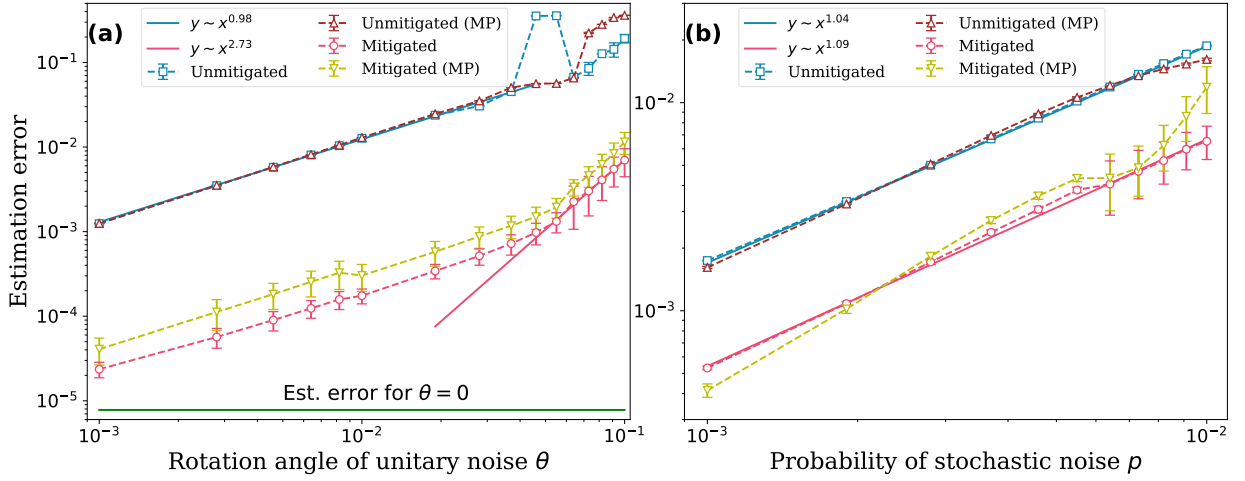

FIG. S4: Simulated results on estimating Floquet quasi-energies with $N_s = 10^7$ shots per circuit and matrix pencil method for data processing. Although the matrix pencil (MP) method can remove the bump in the unmitigated case, the functional fit outperforms MP for most noise parameters.

power of our method). Thus we perform another simulation with $N_s = \infty$ (directly compute expectation values of circuits from their final quantum states) and keep other settings unchanged. In this case, we find there is a scaling almost $\sim \frac{1}{\sqrt{N_r}}$, as shown in Fig. S5(b).

This scaling behavior is explicable. Only when the number of random circuits is infinite can RC transform circuit noise into stochastic Pauli noise exactly. In this case, if the noise channel is denoted in the Pauli operator basis, the off-diagonal elements of noise matrix is zero, which are the expectation values among the noise channels from all random circuits. However, for a finite number of random circuits $N_r$, the average values of off-diagonal elements of noise channels from these $N_r$ random circuits may not be zero, as these average values are random variables with expectation values of 0 and finite statistical deviations. The phase estimation error for $N_r$ random circuits is proportional to the magnitudes of off-diagonal elements (parts of non-Pauli errors), which typically have similar magnitudes to their standard deviations. As the standard deviations decrease with $1/\sqrt{N_r}$, so does the phase estimation error.

### B. Order finding

The order finding problem is to find the least positive integer $r$ such that for two specified coprime numbers $x$ and $N$ ($x < N$) we have $x^r = 1 \pmod{N}$ [15]. This problem is interesting because the integer factorization can be transformed into it. The problem can be solved efficiently by finding a phase of a specific type of unitary operator $U$, which is defined as

$$U|y\rangle = |xy \pmod{N}\rangle \tag{S31}$$

where $|y\rangle$ is a computational basis state. This unitary operator has an invariant subspace that is spanned by $r$ computational states $|x^0 \pmod{N}\rangle, \cdots |x^{r-1} \pmod{N}\rangle$. In this subspace, $U$ has eigenvalue $e^{2\pi i s/r}$ with eigenstate

$$|u_s\rangle = \frac{1}{\sqrt{r}} \sum_{k=0}^{r-1} \exp\left(\frac{-2\pi i s k}{r}\right)|x^k \pmod{N}\rangle \tag{S32}$$

for integers $0 \le s \le r - 1$. One can observe that

$$|\vec{1}\rangle = |0, 0, \cdots, 1\rangle = \frac{1}{\sqrt{r}} \sum_{s=0}^{r-1} |u_s\rangle. \tag{S33}$$

Thus this state $|\vec{1}\rangle$ serves as a good initial state for phase estimation. Here $|\vec{0}\rangle = |0, 0, \cdots, 0\rangle$ is a proper reference state with eigen-phase 0. So, we first prepare the initial state as $|\psi\rangle = \frac{1}{\sqrt{2}}(|\vec{0}\rangle + |\vec{1}\rangle)$ by performing a Hadamard gate on the last qubit and

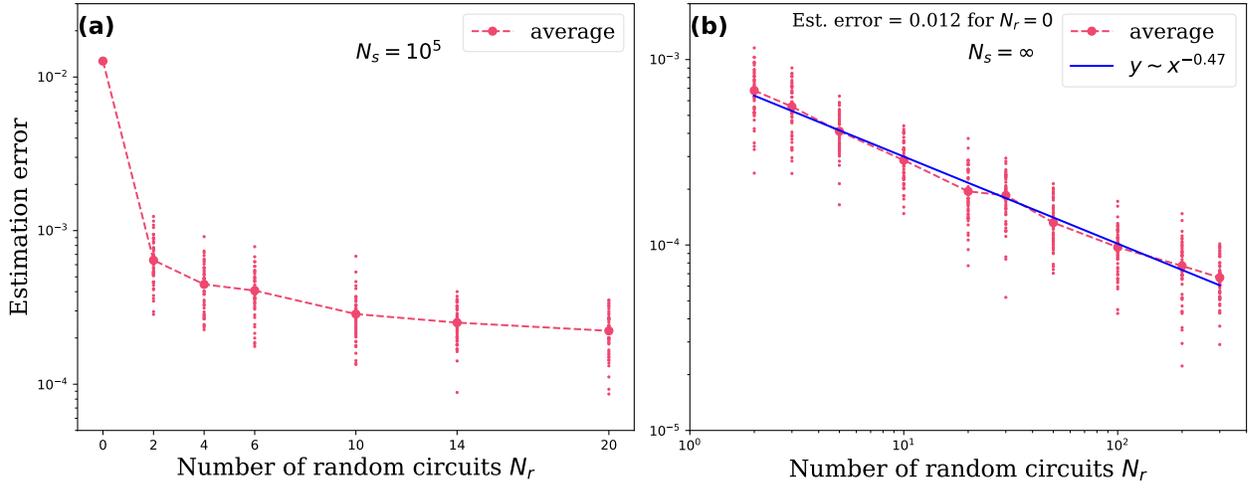

FIG. S5: Estimation error vs. the number of random circuits $N_r$. Each bare circuit is run $N_s = 10^5$ times in (a) and $N_s = \infty$ (directly compute expectation values from quantum states) in (b). For each bare circuit, we generate $N_r$ random circuits in RC and each random circuit is run for $N_s/N_r$ times to keep the resource cost unchanged. $N_r = 0$ means RC is not performed. The results show that the estimation error of phases converges very fast with the number of random circuits in (a) and the converged value is the limit set by the finite number of shots. In (b), we find the phase estimation error scales with $N_r$ almost as $\sim \frac{1}{\sqrt{N_r}}$.

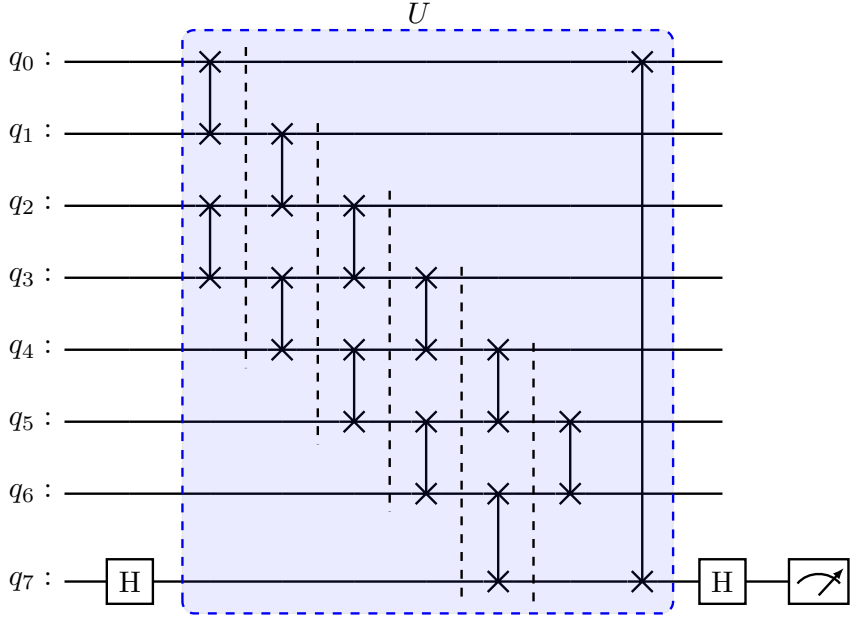

FIG. S6: An example of the circuits used in the simulation of order finding problem with $x = 4, N = 255$. In this example, $L = 1$ and the measurement operator is Pauli-X. The shaded part is the circuit implementing the order finding unitary operator, which contains only SWAP gates.

finally measure the Pauli-X and Pauli-Y of the last qubit. In this work, we study an order finding problem with $x = 4, N = 255$. An example of the circuits used is shown in Fig. S6. We repeat the unitary of interest $U$ up to $L_{\max} = 100$. Other simulation parameters are the same as those in the simulated experiments for estimating Floquet quasi-energies.

We also study the effects of unitary noise and stochastic noise in this problem. The unitary noise added to the single qubit gate is a Pauli-Y rotation error $e^{-i\theta_1/2Y}$ and the two-qubit unitary noise is $e^{-i\theta_2/2Y_1} \otimes e^{-i\theta_2/2Y_2}$. The rotation angle used in

the simulation is $\theta_1 = \theta_2 = 0.1$. For the case of stochastic noise, we apply a $\sqrt{Y} = e^{-i\frac{\pi}{4}Y}$ gate on each qubit with probability $p = 0.05$.

---



\end{document}